# Evidence for transient atmospheres during eruptive outgassing on the Moon

Elishevah M.M.E. van Kooten[1*], Frédéric Moynier[1] and James M.D. Day[2]

[1]Institut de Physique du Globe de Paris, Université de Paris, UMR CNRS 7154, 1 rue Jussieu, 75005, Paris, France

*vankooten@ipgp.fr

[2]Scripps Institution of Oceanography, University of California San Diego, La Jolla, CA 92093-0244, USA

# 1 ABSTRACT

Events following the giant impact formation of the Moon are thought to have led to volatile depletion and concurrent mass-dependent fractionation of the isotopes of moderately volatile elements (MVE). The detailed processes and conditions surrounding this episode remain obscured and are not unified by a single model for all volatile elements and compounds. Using available data, including new Zn isotope data for eight lunar samples, we demonstrate that the isotopic fractionation of MVE in the Moon is best expressed by non-ideal Rayleigh distillation, approaching the fractionation factor α using the reduced masses of the evaporated isotopologues. With these calculations a best fit for the data is obtained when the lunar MVE isotope data is normalized to ordinary or enstatite chondrites ($\Delta_{Moon-OC,EC}$), rather than a bulk silicate Earth (BSE) composition. This analysis further indicates that the parent body from which the Moon formed cannot have partitioned S into its core based on S isotope compositions of lunar rocks. The best fit between $\Delta_{Moon-OC,EC}$ and modelled non-ideal Rayleigh fractionation is defined by a slope that corresponds to a saturation index of 90 ± 4 %. In contrast, the older Highland suite is defined by a saturation index of 75 ± 2 %, suggesting the vapor phase pressure was higher during mare basalt eruptions. This provides the first tangible evidence that the Moon was veiled by a thin atmosphere during mare basalt eruption events spanning at least from 3.8 to 3 billion years ago and implies that MVE isotope fractionation dominantly occurred after the Moon had accreted.

# 2 INTRODUCTION

Apart from the Earth, no other Solar System body can be investigated so thoroughly for its geological evolution as the Moon. The Apollo and Luna sample return missions and complementary lunar meteorites provide significant quantities of material, and remote sensing missions have provided global geological context, in order to study the Moon's dynamic history. Unravelling the detailed formation of the Moon through these sources of information allows insight into the evolution of the Solar System and provides greater understanding of the processes of planet formation within the habitable zone occupied by the terrestrial planets.

One of the key constraints on the formation of the Moon comes from oxygen isotope studies, which show that the Earth and Moon have identical (Greenwood et al. 2018; Hallis et al. 2010; Spicuzza et al. 2007; Wiechert et al. 2001; Young et al. 2016), or at least nearly-identical (Cano et al. 2020; Herwartz et al. 2014) compositions within analytical uncertainty. All other known Solar System bodies show distinct variations in these isotope signatures (Clayton 2003), suggesting that the proto-Earth and the lunar parent body or bodies formed in similar accretion regions (Mastrobuono-Battisti et al. 2015; Wiechert et al. 2001). This assessment is in agreement with identical isotope ratios after correction of mass-dependent isotopic variations for refractory elements in Earth and the Moon (Kruijer & Kleine 2017; Mougel et al. 2018; Schiller et al. 2018; Zhang et al. 2011). However, such a model cannot be reconciled with the Moon's low iron content (Lucey et al. 1995), apparently young – but disputed – formation age (Borg et al. 2011; Connelly & Bizzarro 2016; Halliday 2008; Kleine et al. 2005; Kruijer & Kleine 2017; Thiemens et al. 2019), and the initial angular momentum of the Earth-Moon system (Cameron & Ward 1976). To explain these features, a giant impact between proto-Earth and a hypothetical impactor body was proposed (Cameron & Ward 1976; Hartmann & Davis 1975), which would have ejected Fe-poor material in a protolunar disk surrounding the Earth. An issue with canonical giant impact models is that most of the material that formed the Moon would still have come from the impactor (Canup 2004; Canup & Asphaug 2001). Proposed solutions have been a merger so complete as to erase any pre-existing compositional differences between the two bodies (Canup 2012), an initially fast spinning proto-Earth (Ćuk & Stewart 2012) or turbulent mixing and equilibration in the energetic aftermath of the giant impact (Pahlevan & Stevenson 2007).

Even though the giant impact theory has been widely accepted in a variety of forms, its aftermath is still debated. The nature of the protolunar disk (i.e. vapor, liquid, debris), the condensation and evaporation processes involved between these pre-accretionary media, the chemical interaction between Earth and the Moon and the final differentiation of the Moon (e.g. lunar magma ocean: LMO) are all unresolved topics that provide uncertainties in the interpretation of lunar history. Previous studies of lunar rocks using the mass-dependent isotope composition of moderately volatile elements (MVEs) have shown significant deviations from the bulk silicate Earth (BSE) isotopic composition (Barnes et al. 2019; Boyce et al. 2018; Day et al. 2020a; Herzog et al. 2009; Humayun 1995; Kato et al. 2015; Kato & Moynier 2017; Moynier et al. 2006; Paniello et al. 2012; Pringle & Moynier 2017; Schediwy et al. 2006; Sossi et al. 2018; Vollstaedt et al. 2020; Wang et al. 2019; Wang & Jacobsen 2016; Wing & Farquhar 2015; Wombacher et al. 2008). The quantity of studies examining volatile depletion in planets has risen in the past decade, in part due to advancements in analytical capabilities that have significantly decreased the amount of material required for measurements and increased the reproducibility of the data, allowing resolution of relatively small isotopic differences between the Moon and Earth. The relationship between MVE isotope variations and the observed volatile depletion of the Moon relative to the Earth can help place key constraints on the evaporation and condensation processes within the protolunar disk and during lunar differentiation.

In this work, we include new Zn isotope data for seven magnesian suite (MGS) samples [15455, 76255, 76335, 76535, 77035, 77215, 78328] and one high-Ti mare basalt [10045], which was previously suggested to have anomalous Zn isotope signatures relative to the average composition of lunar basalts (Wimpenny et al. 2018). We then utilize the full data inventory for the mass-dependent isotope fractionation of MVEs collected over the past decades, which we present as a fully accessible data collection within Excel. This collection allows for the cross reference of data between numerous lunar samples. We use this data set to explain the collective behavior of MVEs through the relationship between mass-dependent isotope fractionation with condensation temperature, bond dissociation energy and theoretical isotope fractionation. Furthermore, we highlight the differences between isotope fractionation of mare basalts, potassium rare earth element phosphorous (KREEP) basalts and Highland (MGS, anorthostic) rocks, as well as the gaps in the currently available data set.

# 3 ZINC ISOTOPE DATA OF MGS AND MARE BASALTS

We report the Zn isotope compositions of seven MGS and one mare basalt using the delta notation, which reflects the permil deviation from the JMC Lyon standard (Table 1). For the MGS, portions of 100-300 mg were digested in concentrated HF/HNO$_3$ for 48 hours on a hotplate at 140°C and subsequently dried down and dissolved in aqua regia for 24 hours at the same temperature.

Aliquots of 5 mg were then purified according to small column chemistry procedures described fully by (van Kooten & Moynier 2019).

The MGS measured here have δ$^{66}$Zn compositions between 1.89 ± 0.12 and 7.60 ± 0.12 ‰, in agreement with previous analyses of the MGS suite (Day et al. 2020a; Kato et al. 2015). All data show a mass-dependent relationship (within analytical uncertainty) between δ$^{66}$Zn and δ$^{68}$Zn, reflecting the absence of isobaric interferences. Three of the MGS measured here [15455, 76535, 78235] represent different chips of samples measured previously by Day et al. (Day et al. 2020a). We show here that dissolving and measuring different fractions of the same sample results in similar- but not identical - δ$^{66}$Zn values. For example, the norite 15455 with the highest measured δ$^{66}$Zn value so far, contains fractions with δ$^{66}$Zn values between 7.60 ±0.12 ‰ and 9.27 ±0.12 ‰. The Zn concentrations of the MGS lie between 0.8 and 3.6 ppm and average around 2 ppm. This is indistinguishable from the average of reported mare basalt Zn concentrations (Table S1, see .tar.gz package of supplementary tables).

In contrast, the Zn concentration of the high-Ti mare basalt measured in this study is low (~0.5 ppm) relative to the average for mare basalts (~2 ppm, Table S1, see .tar.gz package of supplementary tables). The δ$^{66}$Zn composition of 10045 is 1.00 ± 0.12 ‰, similar to previous measurements of mare basalts (Table S1, see .tar.gz package of supplementary tables). The mass dependency on 10045 is at lower precision (δ$^{68}$Zn = 2.50 ± 0.24 ‰) due to the very high matrix/Zn ratio of this sample. More refined treatment of further samples with low Zn/Ti ratios will be warranted in the future.

## 4 LIMITATIONS WITHIN THE LUNAR MVE DATASET

In the compiled dataset (Table S1 and references therein, see .tar.gz package of supplementary tables), we consider mass-dependent isotope fractionation of volatile elements between lunar samples and terrestrial standards using the delta notation. Refractory elements (e.g., Ca), usually with 50% condensation temperatures (TC$_{50}$) above that of Li (1150 K), show no significant deviations between the Earth and Moon (Sossi et al. 2016). Hence, we compiled stable isotopic data for elements with TC$_{50}$ < 1000 K (Zn, Cu, Rb, Ga, K, S, Sn, Cd, Se and Cl). Where possible, we have added concentrations of measured elements, as well as the magnesium number (Mg#) of the bulk sample, a proxy of primitive melts or degree of fractional crystallization in igneous rocks. We note that the majority of volatile element isotopic data presently represents bulk sample analyses for lunar rocks. Chlorine isotope data are from lunar apatite grains and melt inclusions (Barnes et al. 2019; Boyce et al. 2015, 2018; Stephant et al. 2019), for which we have used the average apatite composition within each lunar sample. We have included bulk Cl isotope compositions of lunar samples (Table S1, see .tar.gz package of supplementary tables), including their water soluble and structurally bound Cl isotope fractions (Gargano et al. 2020). Apatite grains reflect the largest fraction of Cl in the bulk rock and mass-dependent fractionation between apatite and the melt is suggested to be minor compared to the large isotope fractionations observed for apatite (Boyce et al. 2015). On the other hand, apatite forms at a late stage in most lunar melts, and so may not be a faithful record of magmatic Cl content (Gargano et al. 2020), since the $\delta^{37}Cl$ values of apatites are typically higher than the bulk values. We note that Gargano et al. (2020) find a large range in bulk mare basalts compositions ($\delta^{37}Cl$ ~ 0-12 ‰). The averaged value ($\delta^{37}Cl$ = 4.1 ± 4.0 ‰) is suggested to reflect volatile loss processes from a Giant Impact, whereas the apatite grains record the final degassing of the LMO/local volcanism. It should be noted that the average apatite grain composition of all mare basalts ($\delta^{37}Cl$ = 11.8 ± 1.1 ‰, Table S1, see .tar.gz package of supplementary tables) corresponds to the 'upper limit' bulk mare basalt composition. The bulk value is typically lowered by the water-soluble Cl isotope fraction, which by itself is defined as a deposition of an isotopically light vapor. If we solely consider the structurally bound Cl, the mare basalts are characterized by two populations, one with $\delta^{37}Cl$ similar to the averaged apatites and one with lower values of 2-4 ‰. For one of these samples (15016) Zn isotope data also exists ($\delta^{66}Zn$ = -1.47 ‰), which we omit from the averaged mare basalt Zn isotope composition because it is consistent with containing recondensed isotopically light Zn(Day et al. 2020a). Hence, it is possible that the Cl isotope compositions of bulk mare basalts are similar to averaged apatite compositions, implying that apatites are in fact representative of the bulk.

Chromium isotope data are also compiled, since it has been suggested recently that Cr isotope fractionation between the Moon and Earth is a result of evaporative processes (Sossi et al. 2018). For this reason, we have also compiled data from Mg and Fe isotopes (Poitrasson et al. 2019; Sedaghatpour & Jacobsen 2019; Sossi & Moynier 2017), with similar 50% condensation temperatures as Cr. Finally, we have added Ca isotope data to the excel sheet as a reference for refractory element signatures (Valdes et al. 2014). We note that the 50% condensation temperatures are only relevant for nebular conditions and do not reflect the relative volatility of elements during planetary evaporation, for which the volatility depends among others on temperature, pressure, composition, and oxygen fugacity (Norris & Wood 2017). For example, Cr under planetary accretion conditions is more volatile than Mg and Fe (Sossi et al. 2018).

Lunar samples from the Apollo missions and meteorites span a range of lithologies and compositions. In this dataset, we distinguish ***anorthositic rocks*** (and their ferroan/cataclastic varieties), which are considered to represent the first lunar crust that formed from the LMO by crystallization and flotation of anorthosite. We note that the magnesian anorthosites, which are represented by some lunar meteorites, and are within breccias, are not represented in the MVE isotope dataset. These rocks possibly represent a significant fraction of the lunar crust (Hiesinger & Jaumann 2014). Anorthosites are part of the suite of Highland rocks, which are old (4.56-4.18 Ga) igneous rocks (Warren & Taylor 2015). ***Magnesium suite (MGS) and alkali-suite norites and troctolites*** are also part of the Highlands rocks and are suggested to be slightly younger (Borg et al. 2017), but mostly contemporary plutonic intrusions into the anorthositic crust (Warren, 2003). The MGS likely represent mixtures between melted mafic and crustal cumulates. The ***mare basalts*** (and their high and low-Ti varieties) (Neal & Taylor 1992) are typically significantly younger (3.8 to 3 Ga) than the Highland rocks and are interpreted to represent products of partial melting from mantle cumulates produced during the LMO stage. Although the total volume represented by mare basalts is minor, these basalts are not impact-contaminated (Day & Walker 2015) and are considered to reflect partial melts from a significant fraction of the lunar mantle. KREEP-rich basalts represent low-Ti basaltic melts that have assimilated the largest fraction of late-stage residual melt (i.e. urKREEP) from the LMO stage (Warren & Wasson 1979). Samples that are not discussed here, are ***lunar breccias*** and ***regolith breccias***. These samples are, by their nature, heterogeneous in composition and contaminated and isotopically fractionated by impacts, sputtering from cosmic rays and from solar wind implementation, resulting in isotopic variations that may be unrelated to processes intrinsic to the Moon (Lucey et al. 2006).

The most complete dataset of Lunar MVE is for Zn, which includes numerous samples from the MGS, high and low-Ti basalts, and anorthositic rocks. Data from KREEP-rich samples are limited to a single impact melt breccia sample for Zn (SaU 169), which may not be representative of pure urKREEP (Day et al. 2020a). In general, MVE data for KREEP-rich samples is quite limited. Apart from Cl in apatite, where measurements can be made on polished sections, most elements have one (Ca, Mg, Cr, Zn, Cd) or no measurements (Ga, Sn, Se) for KREEP-rich samples. Most recently, the K isotope compositions of KREEP-rich basalts were measured (Tian et al. 2020). Similar to KREEP, anorthositic rocks are not represented well in the dataset, especially considering the large isotopic variations previously observed for Zn and Ga (Kato et al. 2015; Kato & Moynier 2017). MGS rocks show less isotopic variation than anorthositic rocks, but significantly more than mare basalts, when considering the available Zn isotope data (Day et al. 2020a; Kato et al. 2015)(this study).

The $\delta^{66}$Zn values for MGS rocks vary between 1.9 ± 0.1 and 8.7 ± 0.1 ‰, with a weighted average of 4.6 ± 1.5 ‰ (2SD). This is similar to the highest $\delta^{66}$Zn values measured for the anorthositic rocks ($\delta^{66}$Zn < 4.2 ‰), but distinct from the mare basalts with a weighted average of 1.4 ± 0.1 ‰ (2SD). This average does not include negative values obtained for some mare basalts (5 out of 32, (Day et al. 2020a)), since these values are attributed to mixing and contamination from isotopically light Zn on the lunar surface. A positive correlation between $\delta^{66}$Zn and 1/Zn has previously been proposed, based on a limited number of data (and not in the associated tables), and suggested to be related to isotopic fractionation during evaporative loss during mare basalt eruption (Wimpenny et al. 2018). However, we have analyzed a fraction of one of the high-Ti mare basalts (10045) examined in Wimpenny et al. (2018) for which lower $\delta^{66}$Zn values and higher Zn concentrations were inferred. We propose that the discrepancy between these data is caused by terrestrial contamination, since this would increase the Zn content (<17 ppm) relative to typical mare basalts (0.5-2 ppm) and consequentially lower the $\delta^{66}$Zn values towards terrestrial values. Such contamination could be expected from the purification procedures used by Wimpenny et al. (2018), since Zn was extracted after a Sr purification, thereby increasing the acid and resin blank.

Aside from Zn, the MVE stable isotope data for MGS rocks are more spartan. While Fe, Cr and Zn are well represented (6 to 10 samples), for Rb, Ga and K isotopes, only one MGS sample has been measured (i.e. alkali norite 77215) and for Cu, S, Sn, Se and Cd no data have been reported in the literature. For Mg and Cr isotopes, MGS are seemingly more enriched in the heavy isotopes compared to mare basalts, similar to Zn and Cl, where the difference between these lunar suites is more obvious. The datum for Rb, Ga and K isotope compositions of MGS are so far indistinguishable from mare basalts. The MVE isotope compositions of mare basalts are well-defined and relatively homogeneous compared to MGS and anorthositic rocks (with the exception of Se, (Vollstaedt et al. 2020)). No systematic differences are observed between low-Ti and high-Ti basalts, although for Se the high-Ti basalts may be more fractionated relative to BSE (Vollstaedt et al. 2020). This conclusion is, however, based on only three datapoints. For Cd isotopes, only one datum exists for high-Ti mare basalt 10017, with an isotope signature identical to BSE (Schediwy et al. 2006). It is unlikely that this single data point can be considered representative of the entire mare basalt suite, most especially since 10017 also has an uncharacteristically low $\delta^{66}$Zn value of -5.41 ± 0.04 ‰ (Paniello et al. 2012). On the other hand, Cu, Rb and K isotope signatures for this sample are typical of the mare basalt average composition (Herzog et al. 2009; Pringle & Moynier 2017; Wang & Jacobsen 2016). It is possible that the Cd isotope composition of mare basalts is better represented by a heavier signature observed in KREEP basalt 14310 (Wombacher et al. 2008).

# 5 WHAT DOES LUNAR MVE ISOTOPE FRACTIONATION SIGNIFY?

Numerous physical and mathematical models have been employed to describe the magnitude and direction of MVE depletion and isotope fractionation in the Moon. The most recent models include: 1) Degassing of volatile elements and compounds from the LMO (Day & Moynier 2014; Dhaliwal et al. 2018; Sossi et al. 2018) (**Zn, S, Cl, K, Cr**), 2) post-eruptive degassing of volatiles (Zhang, 2020; **Cu, Zn, Cl, H**), 3) volatile degassing and partitioning into the lunar core (Xia et al. 2019) (**Zn, Cu**), 4) fractionation following the giant impact in the lunar disk (Wang & Jacobsen 2016), 5) incomplete condensation during accretion and subsequent degassing during a LMO phase (Young and Tang, 2019; **K**), 6) Degassing of volatiles into an undersaturated viscous vapor disk (Nie & Dauphas 2019) (**Cu, Zn, K, Rb, Ga**) and 7) Repeated vapor-liquid equilibrium and viscous vapor drainage (Wang et al. 2019) (**Sn, Zn, K**). Most of these models, that were initially proposed by Day and Moynier (Day & Moynier 2014), describe some form of non-ideal Rayleigh distillation. Strictly speaking, ideal Rayleigh distillation implies that the source phase is always perfectly homogenized with an isotopic fractionation occurring under either kinetic or equilibrium conditions:

$$\delta_{final} = \delta_{initial} + [(1000 + \delta_{initial}) \times (f^{\alpha-1} - 1)] \qquad \text{A1}$$

in which $f$ is the fraction of residual volatiles of the Moon relative to the source (i.e. Earth) and $\alpha$ is, often for simplicity, the kinetic fractionation factor:

$$\alpha = \left(\frac{m_x}{m_y}\right)^{\beta} \qquad \text{A2}$$

where $m_x$ and $m_y$ are the average atomic masses of the light and heavy isotopes/isotopologues, respectively. The factor β is usually <0.5 (Richter et al. 2011), when α is experimentally determined by evaporation under vacuum from the slope between -ln(f) and ln(δ). Note that (α-1) x 1000 is also defined as Δkin sometimes (Fig.1; (Nie & Dauphas 2019)).

Day and Moynier (2014) noted that ideal Rayleigh distillation (when α is approached as kinetic fractionation) results in much larger isotope fractionations than observed for lunar samples (e.g. Zn, K, Cl). For example, the calculated $Δ^{66}Zn_{Moon-BSE}$ ($Δ^{66}Zn_{Moon-BSE} = δ^{66}Zn_{Moon} − δ^{66}Zn_{BSE}$) is ~63 ‰ (using α = 0.988, Table S3, see .tar.gz package of supplementary tables; f = 0.007, (Taylor & Wieczorek 2014)), whereas the estimated $Δ^{66}Zn_{Moon-BSE}$ from mare basalts is 1.30 ± 0.13 ‰. It has been suggested that even though there is a large offset between the modelled and the observed isotope fractionation, this offset is systematic and can be defined by a factor given by (1-S), where S is the saturation index of the vapor phase, defined by:

$$S = \frac{P_i}{P_{i,eq}} \qquad \text{A3}$$

Where $P_i$ is the partial vapor pressure of element *i* and $P_{i,eq}$ is its equilibrium vapor pressure at the relevant temperature (Fig. 1,(Nie & Dauphas 2019)). This is one solution to artificially let α approach its value under equilibrium, rather than for purely kinetic isotopic fractionation. Although this method works for some of the MVE isotope data (i.e. Rb, Cu, Ga, K and Zn), Sn, S, Cd, Se and Cl do not lie on such correlations. It should be noted that only the error bars of the observed isotope fractionation are plotted on figures in this manuscript. Errors of the modelled Rayleigh fractionation are much larger (up to 70% of a given value, Table S3, see .tar.gz package of supplementary tables), due to the large uncertainties on the elemental fractions of the Moon relative to the Earth (Table S3, see .tar.gz package of supplementary tables).

Day et al. (2020b) have shown that the fractionation factor for evaporation residues of Zn, Cu and K isotopes in trinitite glasses is higher than for ideal Rayleigh fractionation: α = 0.998-0.999. Although α is usually assumed to be constant to simplify the integration of the equation, in many realistic situations, α can vary during the Rayleigh process (Criss 1999). For large changes in concentration (f < 0.1) the Rayleigh equation is best integrated over a time-dependent distillation trajectory (Criss 1999; Dreybrodt 2016). Rather than α being a constant, it is expected to change with increasing evaporation, temperature and vapor pressure and is, therefore, difficult to predict in natural systems. The modelled isotope fractionation over such a differential distillation trajectory can be approached by using the average α value over the entire range of evaporation. Since it is known from experimental data and analyses of high-temperature nuclear fallout samples and tektites that, during volatile isotope fractionation, α approaches equilibrium fractionation (Day et al. 2017, 2020b; Moynier et al. 2009; Sossi et al. 2020; Wimpenny et al. 2019), the fractionation factor is better estimated using the reduced masses (µ) of isotopologues rather than the molecular or atomic masses:

$$\mu_i = \frac{M \times m_i}{M + m_i} \qquad \qquad A4$$

where $m_i$ is the atomic mass of the isotope species and M is the average atomic mass of the bonding element. Using the reduced mass of an isotopologue considers the bonding energy of molecules (see Appendix A) and, hence, approaches equilibrium fractionation when calculating α. For example, $\alpha_{Zn}$ using the reduced masses of ZnO, ZnS and $ZnCl_2$ (possible bonding of Zn in the melt and vapor phase) are 0.997, 0.996 and 0.995, respectively (Table S3, see .tar.gz package of supplementary tables), which is closer to the empirically obtained values (Day et al., 2020b). Using the α reduced mass for each isotope system and assuming the breaking of bonds of oxides (Zn, Rb, K, Cu, Sn) and sulfides (Ga, Cd, S) during evaporation, we obtain a good fit ($R^2$ =0.97, Fig. 2A) between the modelled and observed fractionation for all elements. Note that the calculated $\Delta Sn_{Moon-BSE}$ approaches zero, more consistent with observed negative $\Delta Sn_{Moon-BSE}$ (Wang et al. 2019), because the fraction of elemental Sn for the Moon/BSE is approximately one (Jochum et al. 1993; Wang et al. 2019). This fit would be even better if we apply the empirically derived α values for Zn, Cu and K (Day et al. 2020b). In particular, the calculated $\alpha_K$ (0.994) is much lower than the experimental $\alpha_K$ (0.998), for which the latter results in a modelled Rayleigh fractionation that is an order of magnitude lower (Table S3, see .tar.gz package of supplementary tables). For most elements like Zn, displaying both chalcophile and lithophile behavior (e.g., Cu, Sn, Ga, Cd), the calculation of the oxide or sulfide isotopologue α values, does not significantly change the fit in Figure 2A. For example, calculating $\alpha_{Cu}$ as CuS or CuO results in Δ values of 11 and 7, respectively.

The matter is different for S: whether $\alpha_S$ is calculated for $S_2$, $SO_2$, COS or $H_2S$ (Wing & Farquhar 2015; Zheng 1990) has a large effect on the Δ value. $SO_2$, $S_2$ and COS have significantly lower $\alpha_S$ (0.985-0.986) than $H_2S$ (0.998), resulting in a far lower Δ value for the latter. The degassing of S as sulphate or sulfide is dependent on the oxygen fugacity of the melt. When outgassing $SO_2$, to obtain an enrichment of $δ^{34}S$ in the melt, the fraction of sulfide/sulphate in the melt should be low, suggesting oxidizing conditions with possibly low $H_2O$ contents. To obtain the same isotope enrichment, when outgassing $H_2S$, the conditions should be more reducing (Zheng 1990). The modelled $fO_2$ conditions of the lunar magma ocean suggest that S was mainly present as $H_2S$ (Deng et al. 2020; Gaillard et al. 2015). Under these conditions, when modelling Rayleigh fractionation using BSE as the parental source for the Moon, then the observed S isotope fractionation does not follow the trend of Fig. 2A and assuming all MVE evaporated under the same saturation conditions. If we model Rayleigh fractionation with a parental source similar to ordinary or enstatite chondrites, the observed sulfur isotope fractionation matches the modelled fractionation better and the entire fit for all elements is improved relative to Fig. 2A ($R^2$ = 0.99, Fig. 3). This implies that the parental source is either a chondritic body or a differentiated body with a sulfur-poor core, in agreement with giant impact models that suggest most of the Moon came from the impactor. Alternatively, the S isotope signature of the Moon is more reflective of the BSE at the time of the giant impact, if after this event effects of core formation on the S isotope composition were significant (Labidi et al. 2013). Nevertheless, without knowing the appropriate fractionation factor of S during evaporation, these interpretations remain speculative. The outline of MVE stable isotopes treated as a collective dataset, highlights the need for experimentally and empirically obtained fractionation factors relevant to volatile evaporative loss (Day et al. 2017, 2020b; Sossi et al. 2020; Wimpenny et al. 2019) and for better constraints on the elemental Moon/BSE fractions (see Appendix A).

If we interpret the slope between the observed and modelled fractionation from Figs. 2A and 3 in a similar manner as Nie and Dauphas (2019), we obtain a saturation index of S=89±2% and 92±2%, respectively. The errors are derived from the slope in Figs. 2A and 3, which was obtained using Isoplot Yorkfit and incorporating 2σ errors of the input data. This implies a less saturated vapor layer within the protolunar disk than found by Nie and Dauphas (2019), resulting in higher estimations of the disk's kinematic viscosity (Fig. 2B, see Table S4 for equations derived from (Nie & Dauphas 2019), see .tar.gz package of supplementary tables). However, with low γ values (the evaporation coefficient) of K, a lower saturation index is still in agreement with the predicted α (viscosity) values of astrophysical models (Gammie et al. 2016).

# 6 TRANSIENT ATMOSPHERES ON THE MOON

The isotopic compositions of mare basalts are typically used to model volatile related fractionation processes, because these samples are thought to be most representative of their parent melts (Warren & Taylor 2015), although not necessarily of the bulk Moon (Sossi & Moynier 2017). For example, the anorthositic rocks show both significant enrichments and depletions in $\delta^{66}Zn$ (Kato et al. 2015). This is likely the result of an initially heavy Zn isotope signature for MVE that was later affected by recondensation of isotopically light Zn during crust formation. The highest $\delta^{66}Zn$ value found in anorthositic rocks ($\delta^{66}Zn$ =4.24±0.04 ‰) is identical to the average MGS composition ($\delta^{66}Zn$ =4.3±1.5 ‰, this study,(Day et al. 2020a)), in agreement with a common process for the formation of MGS and anorthosites. Along with Zn, Cl isotope compositions are also enriched in MGS. It has been suggested that the MGS are the product of mixing with a Cl-rich and $^{37}Cl$-rich endmember and that a correlation exists between Cl concentration and $\delta^{37}Cl$ values in apatite grains from mare basalts, KREEP and MGS (Boyce et al. 2015) (Fig. 4). However, additional Cl isotope data could not verify such a correlation and imply that trends between Cl and $\delta^{37}Cl$ possibly only exist for low-Ti and KREEP basalts (Barnes et al. 2019; Boyce et al. 2018; Potts et al. 2018). While Cl concentrations and $\delta^{37}Cl$ values for MGS apatite grains are indeed higher than basalts (only two samples measured so far for MGS: 76535 and 78235), there is no continuous correlation between basalts, KREEP and MGS. Each suite has fairly constant $\delta^{37}Cl$ values over a range of Cl concentrations. The two basalts with higher $\delta^{37}Cl$ values and Cl concentrations have large uncertainties (Potts et al. 2018) and, as discussed above, sample 10017 is likely not representative of the basalt suite. In addition, no correlation is observed between the Zn content and $\delta^{66}Zn$ values (Fig. 5), implying that MGS were not formed by mixing of reservoirs with various Zn isotope compositions, but rather through a common MVE isotope fractionation process such as evaporative loss.

Modelling of the Zn isotope evolution during LMO crystallization shows that early-formed olivine and Zn-rich cumulates have lower $\delta^{66}Zn$ values than later-formed sequences, since these cumulates would have had longer time to degas and, hence, fractionate Zn to a greater extent (Dhaliwal et al. 2018). According to this model, KREEP basalts are expected to have the heaviest Zn isotope signatures, since they represent the last dredges of magma ocean crystallization. In contrast, recent data that show identical isotope signatures between mare and KREEP-rich basaltic meteorite (Day et al. 2020a).

This could be explained by mixing between KREEP and mafic components, increasing the Mg# and lowering the $\delta^{66}$Zn, but it could also indicate that there was no change in $\delta^{66}$Zn during magma ocean processes. For example, models with an early stagnant lid on top of the magma ocean would prevent degassing and isotope fractionation over time (Dhaliwal et al. 2018). Furthermore, no significant variations are observed between high-Ti and low-Ti mare basalts (Table S1, see .tar.gz package of supplementary tables), even though they are interpreted to be derived from different sources (Snyder et al. 1992) and at different times (Snyder et al. 1994). In an isotopically stratified LMO, different source materials would likely reflect different isotopic signatures in the magmatic products.

Here, we provide an alternative explanation for the observed MVE stable isotope differences between MGS and mare basalts. First, we exploit the common spatiotemporal origin of ferroan anorthosite and magnesian suite rocks. Considering anorthosites and MGS formed virtually contemporary and at/near the lunar surface, we should expect the MVE isotope systematics related to volatile loss to reflect the same. As discussed above, the average Zn isotope composition of MGS corresponds to the upper limit value of the anorthosites, which we assume is not contaminated by condensation or mixing (Day et al. 2020a). If we examine the limited data available for MGS accordingly and include the heaviest isotope compositions of anorthosites (i.e. Ga and Cd), then it becomes apparent that the ferroan anorthosites and MGS are enriched in heavier isotopes for the most volatile elements (> Rb, K) relative to mare basalts. For example, the Rb and K isotope composition of MGS is so far identical to mare basalts (Humayun 1995; Pringle & Moynier 2017; Tian et al. 2020), but Zn, Cd and Cl show heavier signatures (Barnes et al. 2019; Day et al. 2020a; Schediwy et al. 2006). If we place this into context with a slope in Fig. 3 equaling the saturation index (Nie & Dauphas 2019), then the saturation of the vapor phase was 76 ±2 % when MGS and anorthosites formed, which is significantly lower than mare basalts (S = 89-92 ±2 %). This implies that the first lunar differentiation event that formed the ferroan anorthosite and MGS was degassed under more undersaturated conditions than the source rock of the mare basalts (Fig. 6). In other words, a higher vapor pressure of all elements is needed to explain the degassing process of the mare basalts. The underlying assumption of this model is that the LMO had a homogeneous MVE isotope composition and was not isotopically stratified (Dhaliwal et al. 2018). If we place this into context with Zn isotopes, the assumption would be that the crystallized LMO had a homogeneous $\delta^{66}$Zn value similar to mare basalts (<1.3 ‰), resulting from evaporation and isotope fractionation during the Giant Impact and/or molten LMO stage (Fig. 6A). During the crystallization of the LMO, the anorthosites and the intrusive MGS at the lunar surface can lose volatiles in a process called 'excess degassing', where the amount of gas released from an unerupted magma exceeds the erupted counterparts (Shinohara 2008; Wallace 2001). If this degassing happens under vacuum conditions, it leads to a more fractionated MVE isotope composition relative to the initial LMO signature ($\delta^{66}$Zn ≈ 4 ‰, Fig. 6B). Since the mare basalts have a less fractionated MVE isotope composition, we propose that they erupted under higher vapor pressure and experienced limited isotope fractionation relative to the anorthosites and MGS (Fig. 6C). This is reconciled with the existence of one or more thin atmospheres formed during mare basalt eruptions (Needham & Kring 2017). This atmosphere was modelled to exist for approximately 70 Ma and could have reached an atmospheric pressure of about 1 kPa. Hence, the MVE isotope variations between the anorthosites/MGS and mare basalts may provide a primary line of evidence for the existence of such an atmosphere. The thin and patchy atmosphere observed on Jupiter's moon Io is related to volcanic degassing and may be a proxy to our Moon's eruptive history (Moses et al. 2002). Moreover, these results indicate that MVE isotope variations in lunar samples cannot be interpreted solely as volatile loss processes from the protolunar disk or from the solar nebula, but should be related to degassing from differentiation events on an already accreted Moon.

We summarize that although the data at hand points towards the existence of early transient atmospheres on the Moon at the time of mare basalt eruptions, the MVE isotope dataset also highlights the limitations of our model. Filling the gaps in this dataset, particularly within the anorthosites, MGS and KREEP samples, are required to improve current interpretations and models on lunar volatile loss.

# 7 ACKNOWLEDGEMENTS

We thank two anonymous reviewers for their helpful comments that have strengthened the paper. This project has received funding from the European Union's Horizon 2020 research and innovation programme under the Marie Skłodowska-Curie Grant Agreement No 786081. F.M. acknowledges funding from the European Research Council under the H2020 framework program/ERC grant agreement no. 637503 (Pristine) and financial support of the UnivEarthS Labex program at Sorbonne Paris Cité (ANR-10-LABX- 0023 and ANR-11-IDEX-0005-02). Parts of this work were supported by IPGP multidisciplinary PARI program, and by Region Île-de-France SESAME Grant no. 12015908.

# 8 APPENDIX A:

In addition to the fractionation factor, the modelled Rayleigh fractionation (Fig. 2A, Table S3, see .tar.gz package of supplementary tables) is also dependent on the fraction $f$ of an element concentration in the Moon over an element concentration in the source body. The elemental abundance of the bulk Moon and BSE are modelled estimations and, hence, have large uncertainties (~50%). As such, even though the modelled Rayleigh fractionation (Fig. 2A and 3) provides a reasonably good fit to the observed fractionation, the scatter in the trend does not allow for a definitive determination of the source body from which the Moon was derived or predictions for the isotope compositions of other underrepresented volatile elements measured in lunar samples (i.e. Cd, Se, Hg). However, $f_{Moon/BSE}$ displays a good correlation with the bond dissociation energy (BDE) of elemental bonds (Fig. S1-A, Albarède et al., 2015), more so than with the 50% condensation temperature as has been previously pointed out (Albarède et al. 2015; Norris & Wood 2017; Sossi et al. 2019; Xia et al. 2019). The slope of this correlation (BDE = 190* $f_{Moon/BSE}$) can be implemented in the general relation between Rayleigh isotope fractionation and $f_{Moon/BSE}$:

$$\Delta = -ln(f_{Moon/BSE} \times 190)$$

Which results in the predicted correlation between $\Delta_{Moon-i}$ and the BDE (Fig. S1-B and S1-C). Here, $\Delta_{Moon-i}$ is normalized to $(m_2-m_1)/m_1*m_2$ (Table S2, see .tar.gz package of supplementary tables), where $m_1$ is the atomic mass of the light isotope and m2 that of the heavy isotope (i.e. $m_2-m_1$ = 2 for $\delta^{66}Zn$). This normalization is done to correct for the increase in isotope fractionation with a larger mass difference and decreasing atomic mass (Schauble 2011). A better fit is obtained plotting against a chondritic source body, rather than BSE. As mentioned above, the BDE of sulfides are typically within range of oxides (i.e. Zn, Cd, Cu, Sn, Se), so it does not matter if they behaved as chalcophile or lithophile species during evaporation from the lunar mantle. For Ga, the BDE of Ga-O is similar to Ga-Te (Luo 2007), which is structurally similar to S-Ga, for which the BDE is unknown. The BDE of Cl is highly variable, depending on its bonding between other halogens, non-metals and transition metals such as Zn (e.g. low BDE) or metals and Cl-H (e.g. high BDE). Chlorine has been proposed to degas as HCl with relatively low Cl/H ratios, but with increasing evaporation degasses as metal halide (i.e. $FeCl_2$ and $ZnCl_2$), when H is depleted in the lunar magma (Sharp et al. 2013; Ustunisik et al. 2015). The evaporation of $ZnCl_2$, for example, could account for the large isotopic fractionations observed for Cl. For Cd, a terrestrial isotope composition (Schediwy et al. 2006) does not fit with the predicted correlation between BDE and $\Delta_{Moon-I}$ (Fig. 4B and 4C), but the more fractionated KREEP basalt value does fall within the predicted range. For Se, the high BDE of Se-S and S-O (Luo 2007) suggests that the Se isotope signature of the lunar mare basalts is terrestrial, like the value obtained for low-Ti basalt 12021 (Vollstaedt et al. 2020). However, the high BDE of Se-S and Se-O is inconsistent with the large elemental fractionation observed between ordinary/enstatite chondrites and lunar samples ($f_{Moon/EC}$ = 0.009, Table S3, see .tar.gz package of supplementary tables). Hence, if the source body where OC- or EC-like, a larger isotopic fractionation would be expected than for a BSE source, where $f_{Moon/BSE} > 1$.

**Table and Figure captions:**

Table 1: Zn isotope data of MGS and mare basalts. Measurements are from this study, Kato et al. (2015) and Day et al. (Day et al. 2020a).

Figure 1: The modelled ideal Rayleigh fractionation Δkin*ln(f) plotted against the observed fractionation (adapted from (Nie & Dauphas 2019)) between lunar basalts and BSE. Elements used by Nie and Dauphas (Nie & Dauphas 2019) are plotted as black spheres, whereas additional data is in blue. The Cd isotope data is uncertain, but both datapoints are shown.

Figure 2: A) The calculated Rayleigh fractionation between lunar basalts and Earth (see Table S3 and text for details) and the observed isotope fractionation. Black dots are for an α value using molecular masses and blue are for $α_{rm}$ values using reduced masses. Dashed lines are the linear trends plotted for both datasets and with $α_S$ calculated as $H_2S$; $f_{Moon/BSE}$ for Se is assumed to be approximately 1 (Vollstaedt et al. 2020), similar to Sn, and the $Δ_{Moon-BSE}$ is the upper limit value. The $α_{Cl}$ is calculated using the reduced mass of $ZnCl_2$. Solid blue lines represent the error envelope of the blue trend. Note that the x-axis is shown as a logarithmic scale for a better display of the data range. The errors of the modelled Rayleigh fractionation are not displayed for better visualization and are shown in Table S3 (see .tar.gz package of supplementary tables). B) The temperature plotted against the kinetic viscosity of the protolunar disk, calculated using the saturation index of K calculated from figure 2A (S = 89%) and Nie and Dauphas (Nie & Dauphas 2019) (S = 99%) and the range of γ values (the evaporation coefficient) found for K. γ is experimentally determined to be 0.13 in a vacuum and 0.017 with a $H_2$ pressure of 9 × $10^{-5}$ bar (Fedkin et al. 2006).

Figure 3: The calculated Rayleigh fractionation between lunar basalts and ordinary chondrites (green OC) or enstatite chondrites (grey EC) and the observed isotope fractionation (see Table S2 and text for details). The Δ value of sulfur is calculated using $α_{H2S}$. Note that the x-axis is shown as a logarithmic scale for a better display of the data range. The errors of the modelled Rayleigh fractionation are not displayed for better visualization and are shown in Table S3 (see .tar.gz package of supplementary tables).

Figure 4: The Cl contents versus $δ^{37}Cl$ values of averaged lunar apatites and melt inclusions. Data are from (Sharp et al. 2010), (Barnes et al. 2019), (Boyce et al. 2015), (Potts et al. 2018) and (Stephant et al. 2019). Basalts with blue error bars are from (Potts et al. 2018). Dashed lines reflect the average composition of each suite. Note that mare basalts are typically grouped into two different chemical suites (e.g., low-Ti and high-Ti), but are indistinguishable in Cl isotope space.

Figure 5: Zinc contents versus the $\delta^{66}$Zn values of lunar basalts (blue spheres), MGS (yellow diamonds) and anorthositic rocks (brown diamonds). Dashed lines represent averaged compositions of each suite. The grey arrows represent increases in Zn content and corresponding decreases in $\delta^{66}$Zn values through mixing and condensation of isotopically light Zn to the lunar surface.

Figure 6: Schematic representation of lunar planetary processes. See main text for more details.

Figure S1: A) Correlation between the $f_{Moon/BSE}$ (Chen et al. 2015; Grewal et al. 2019; Jochum et al. 1993; Khan et al. 2006; McDonough & Sun 1995; Righter 2019; Ringwood 1992; Sossi et al. 2018; Taylor & Wieczorek 2014), the 50% condensation temperature (Wood et al. 2019) and the BDE (Luo 2007). B and C) The normalized lunar isotope signatures against BDE of OC and BSE.

| Sample # | Section # | Type | $\delta^{66}Zn$ | 2SD | $\delta^{68}Zn$ | 2SD | Zn (ppm) | n | Reference |
|---|---|---|---|---|---|---|---|---|---|
| 15445 | 333 | Norite | 2.46 | 0.29 | 5.40 | 0.44 | 0.8 | 2 | Day20 |
| 15455 | 391 | Norite | 9.27 | 0.11 | 18.82 | 0.21 | 2.0 | 2 | Day20 |
|  | 400 | Norite | 7.60 | 0.12 | 15.46 | 0.24 | 2.0 | 1 | this study |
| 72415 |  | Dunite | 6.27 | 0.04 | 12.43 | 0.08 | 2.6 | 1 | Kato17 |
| 76255 | 137 | Norite | 1.89 | 0.12 | 3.40 | 0.24 | 3.3 | 1 | this study |
| 76335 | 83 | Troctolite | 2.37 | 0.12 | 5.38 | 0.24 | 1.5 | 1 | this study |
| 76535 | 186 | Troctolite | 3.14 | 0.12 | 6.58 | 0.24 | 1.5 | 1 | Day20 |
|  | 196 | Troctolite | 2.52 | 0.12 | 8.42 | 0.24 | 1.5 | 1 | this study |
| 77035 | 269 | Norite | 3.54 | 0.12 | 7.09 | 0.24 | 2.9 | 1 | this study |
| 77215 |  | Alkali norite | 3.04 | 0.12 | 5.95 | 0.24 | 3.6 | 1 | Kato17 |
| 78235 | 156 | Norite | 3.50 | 0.28 | 7.10 | 0.56 | 2.0 | 2 | Day20 |
|  | 172 | Norite | 3.35 | 0.12 | 6.99 | 0.24 | 2.0 | 1 | this study |
| 78328 |  | Norite | 2.05 | 0.12 | 6.82 | 0.24 | 2.0 | 1 | this study |
| 10045 |  | High-Ti | 1.00 | 0.12 | 2.50 | 0.24 | 0.5 | 1 | this study |

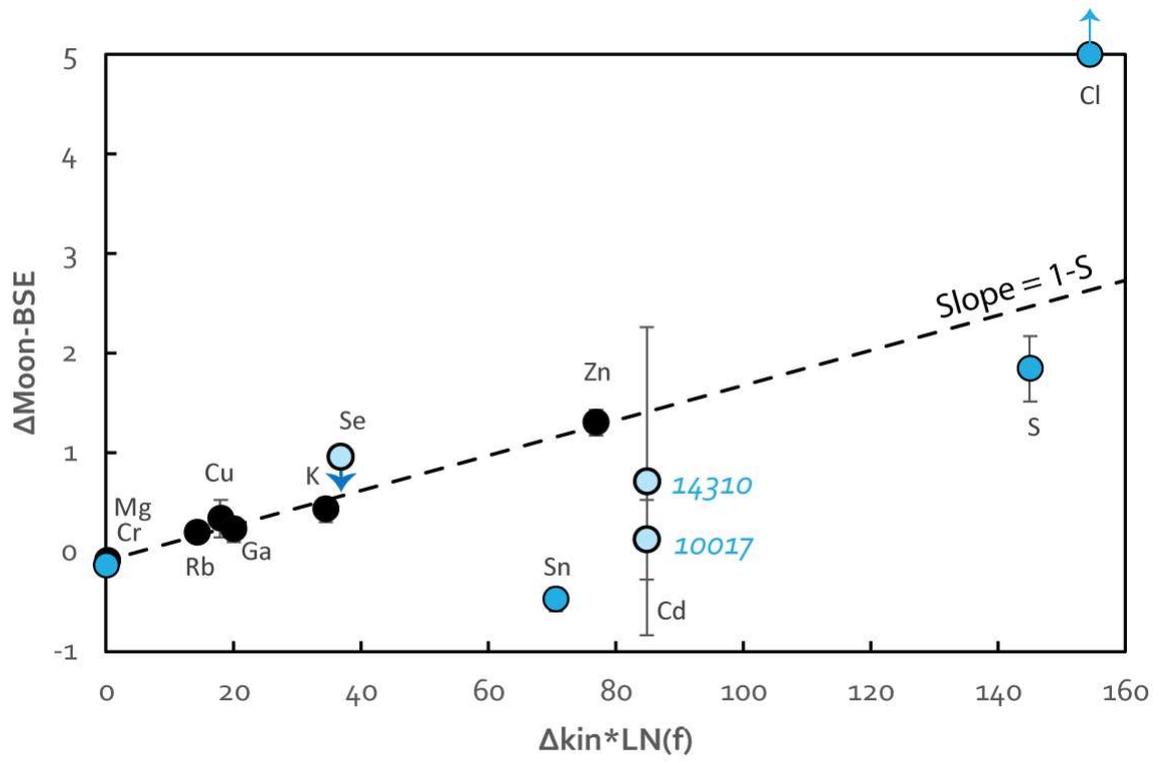

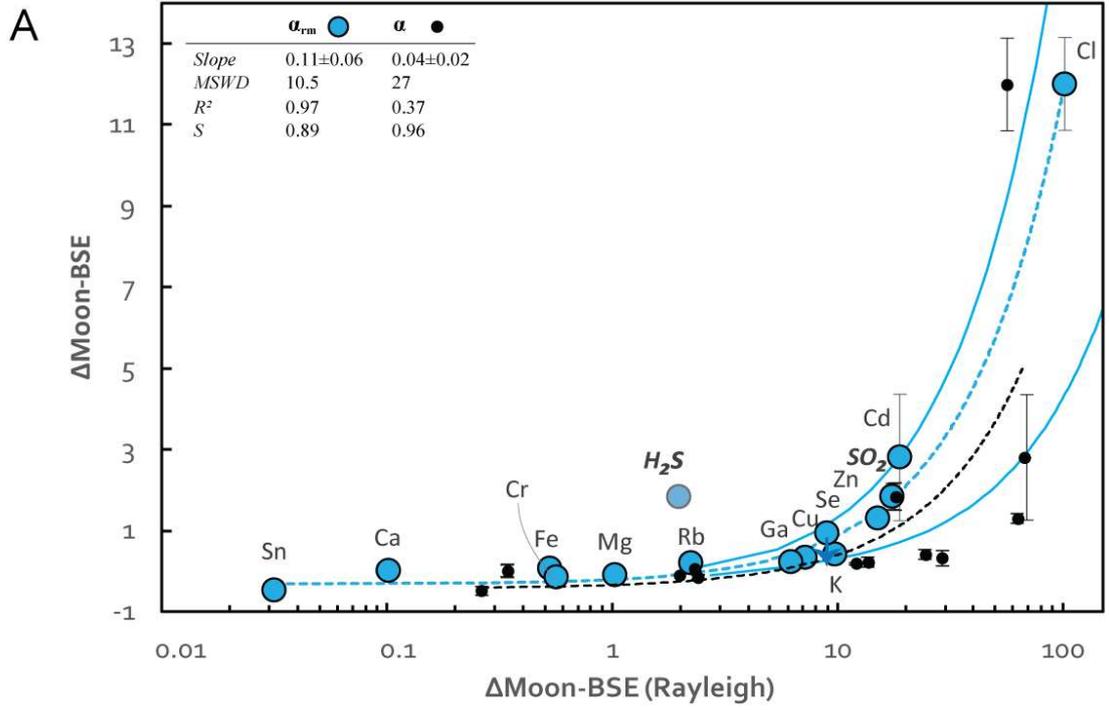

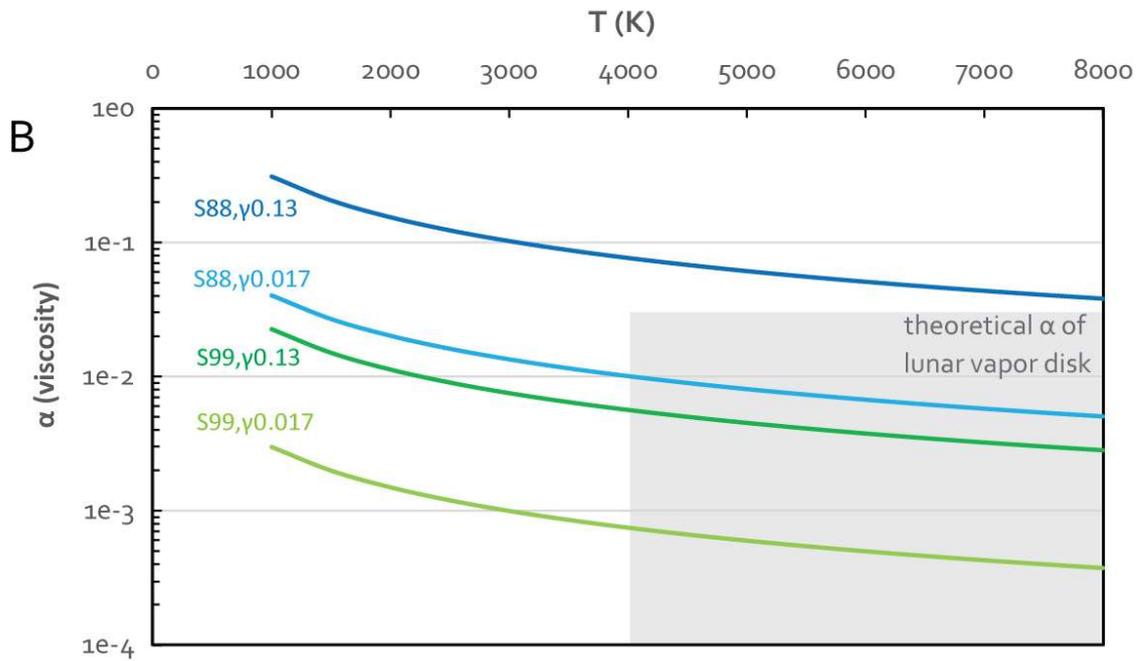

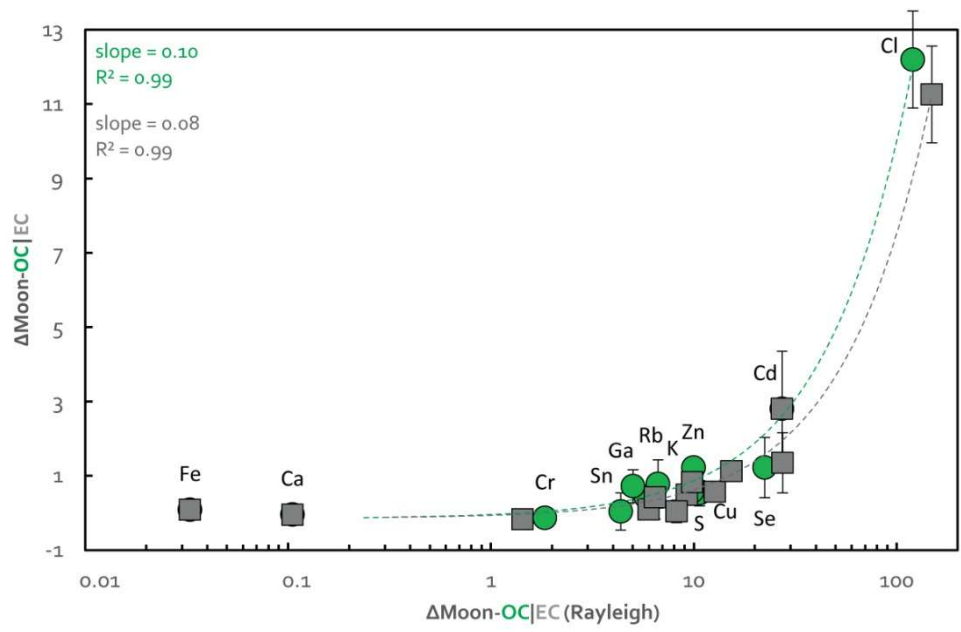

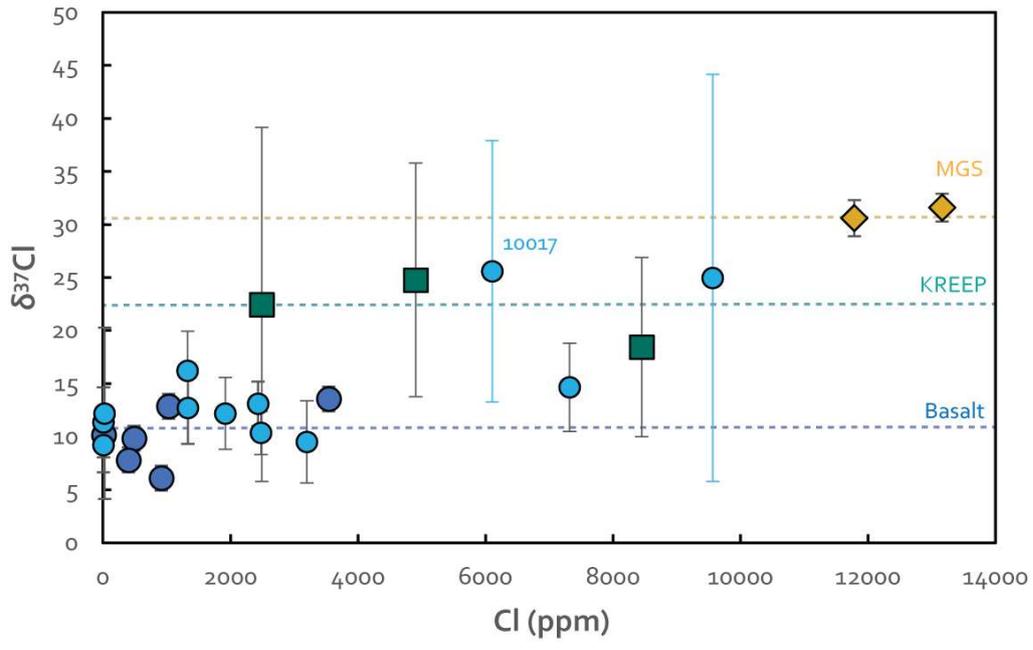

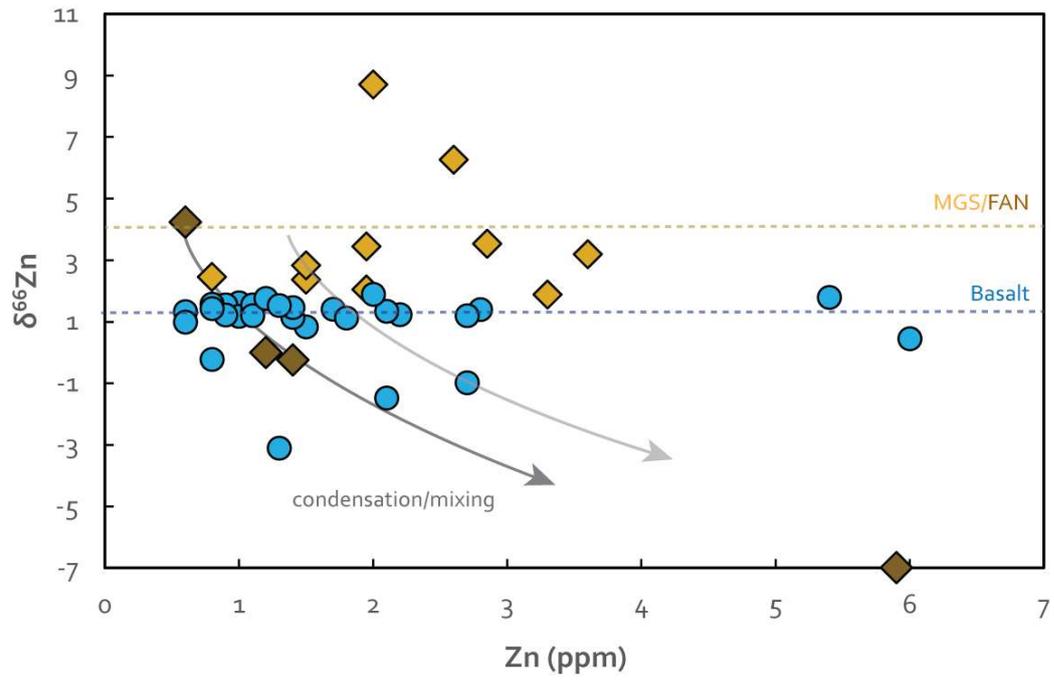

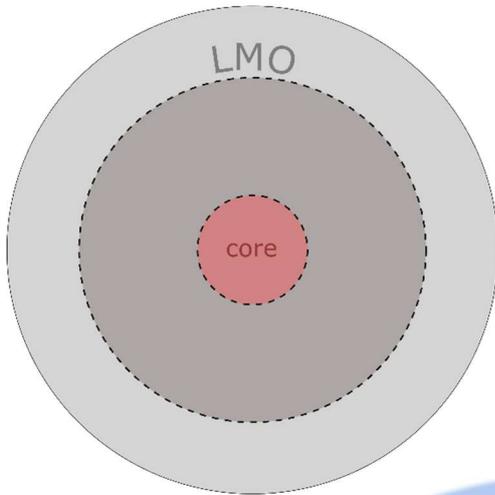

**A** 50-200 Myr after $t_0$: Giant Impact and/or LMO degassing results in MVE depletion and isotope fractionation.

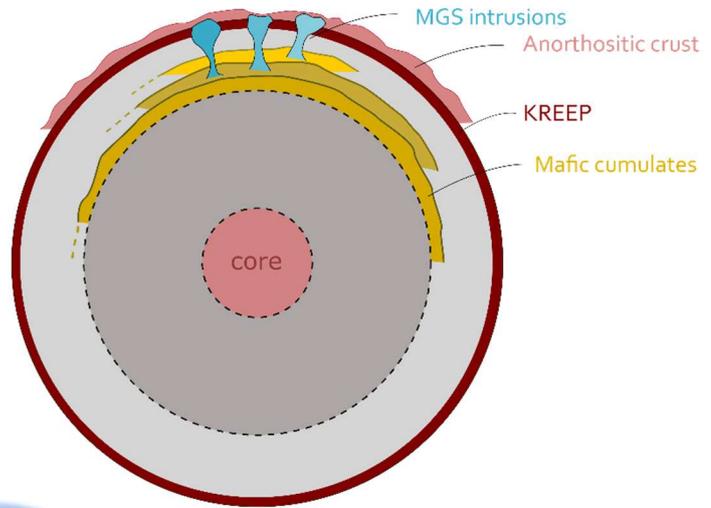

**B** 4.5-4.2 Ga: Crystallization of the LMO, degassing of anorthositic crust and intruding MGS under vacuum

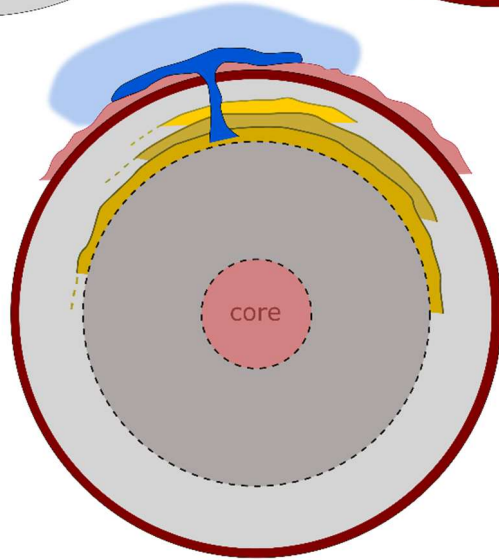

**C** 3.8 - 3.0 Ga: Eruptions of mare basalts under saturation of transient atmospheres.

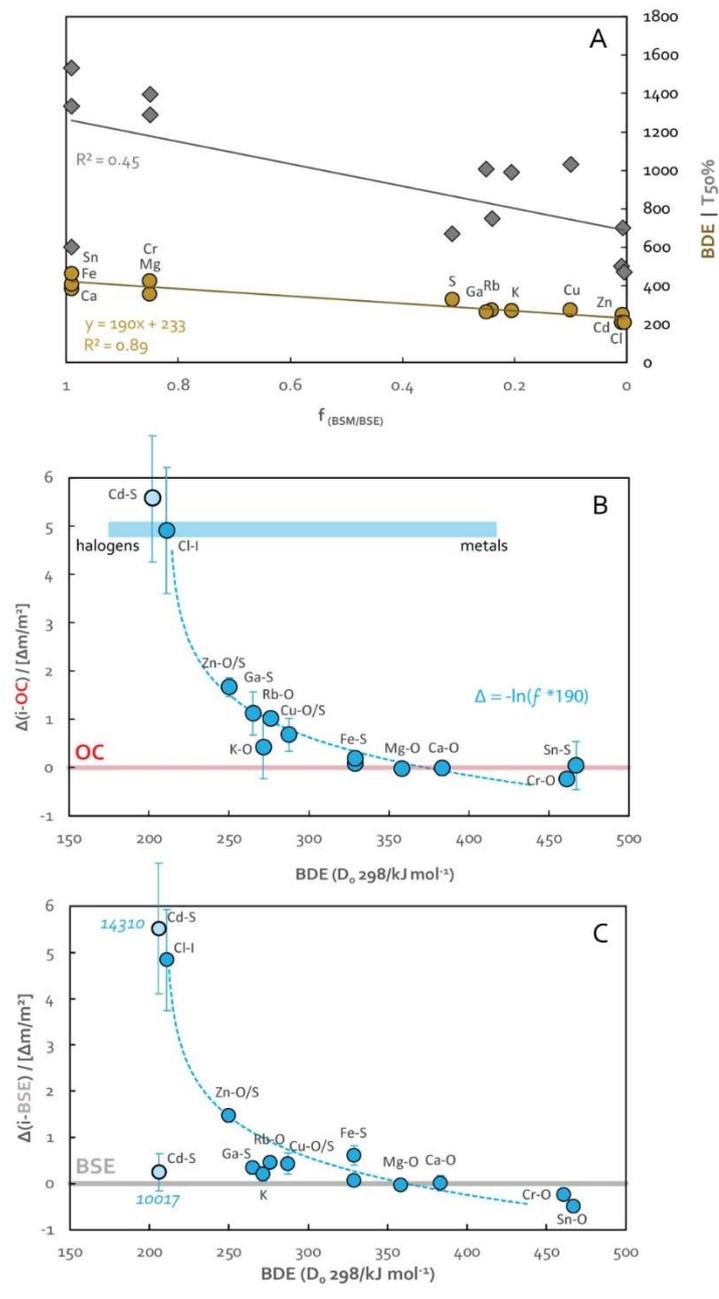